\documentclass[runningheads,a4paper,orivec]{llncs}

\usepackage[dvips]{graphicx}
\usepackage{subfig,multirow,rotating}
\usepackage[left]{lineno}
\usepackage{tabularx,multirow}
\usepackage{icomma,units}
\usepackage{enumerate}
\usepackage{booktabs}
\usepackage{graphicx,subfig}
\usepackage{amsmath,amsfonts,amssymb}

\usepackage{listings}
\usepackage[dvipsnames,usenames]{xcolor}
\usepackage{paralist}
\usepackage{amsmath}
\usepackage{amssymb}
\usepackage{dsfont}
\usepackage{tabulary}
\usepackage{MnSymbol}
\usepackage{stmaryrd}
\usepackage{graphicx}

\usepackage{fancyhdr}

\usepackage{url}

\usepackage{xspace}
\usepackage[nonumberlist,acronym,toc,numberedsection=autolabel]{glossaries}
\usepackage[colorinlistoftodos,color=orange,textsize=tiny]{todonotes}
\usepackage{algorithm}
\usepackage{algorithmic}

\newcommand{\CASE}[1]{\STATE \textbf{case} #1\textbf{:} \begin{ALC@g}}
\newcommand{\ENDCASE}{\end{ALC@g}}

\newcommand{\DEFAULT}{\STATE \textbf{default:} \begin{ALC@g}}
\newcommand{\ENDDEFAULT}{\end{ALC@g}}
\newcommand{\DEFAULTLINE}[1]{\STATE \textbf{default:} }

\usepackage{lscape}

\urldef{\mailBT}\path|{stefan.jablonski}@uni-bayreuth.de|
\urldef{\mailVie}\path|{stefan.schoenig}@wu.ac.at|

\newfloat{lang}{thp}{lop}
\floatname{lang}{Language}

\begin{document}
\lstset{
  breaklines=true,                 
  keywordstyle=\color{blue},       
  language=SQL,                 
  numbers=left,                    
  numbersep=15pt,                   
  rulecolor=\color{black},         
  stepnumber=1,                    
  tabsize=2,	                   
}

\title{SQL Queries for Declarative Process Mining on Event Logs of Relational Databases\thanks{This work was funded by the European Union’s Seventh Framework Programme (FP7/2007-2013) grant 612052 (SERAMIS). }}

\titlerunning{SQL Queries for Declarative Process Mining}

\author{Stefan Sch\"onig}

\authorrunning{Sch\"onig S.}

\institute{Vienna University of Economics and Business, Austria\\ \mailVie}

\maketitle
\thispagestyle{empty}

\begin{abstract}
Flexible business processes can often be modelled more easily using a declarative rather than a procedural modelling approach. Process mining aims at automating the discovery of business process models. Existing declarative process mining approaches either suffer performance issues with real-life event logs or limit their expressiveness to a specific set of constaint types. Lately, with RelationalXES a relational database architecture for storing event log data has been introduced. In this technical report, we introduce a mining approach that directly works on relational event data by querying the log with conventional SQL. We provide a list of SQL queries for discovering a set of commonly used and mined process constraints. 


\end{abstract}

\section{Introduction} \label{sec:introduction}
Process mining is the area of research that embraces the automated discovery, conformance checking and enhancement of business process models. All involved techniques are evidence-based, as the input is always event logs that comprise a collection of computer-recorded information tracking the executions of process instances.

Two different types of processes can be distinguished \cite{Jablonski1994}: well-structured routine processes with exactly predescribed control flow and flexible processes whose control flow evolves at run time without being fully predefined a priori. Flexible processes are common in healthcare, where for example patient diagnosis and treatment processes require flexibility to cope with unanticipated circumstances. In a similar way, two different representational paradigms can be distinguished: procedural models describe which activities can be executed next and declarative models define execution constraints that the process has to satisfy. The more constraints we add to the model, the less possible execution alternatives remain. As flexible processes may not be completely known a priori, they can often be captured more easily using a declarative rather than a procedural modelling approach \cite{VanderAalst2009,Pichler2012,Vaculin2011}.

Declarative languages like \emph{Declare} \cite{pesic2006declarative}, \emph{Dynamic Condition Response} (DCR) graphs \cite{hildebrandt2013contracts} or \emph{Declarative Process Intermediate Language} (DPIL) \cite{zeising2014towards} can be used to represent these models, and tools like DeclareMiner \cite{Maggi2011}, MINERful \cite{di2013two} or DPILMiner \cite{schoenig2015bpmds} offer the capabilities to discover such models automatically from event logs. Existing declarative process mining approaches either suffer from performance issues with real-life event logs or limit their search space to a specific and fixed set of constaints in order to be able to cope with the size of real-life event logs. To the best of our knowledge an approach that is fast and customisable does not exist.

We fill this research gap by introducing a declarative mining approach that works on event data that is stored in relational databases by querying the log with conventional SQL. Process mining by means of SQL queries turns out to be an integrated and language overspanning solution to process discovery.

In this technical report we provide a list of SQL queries for discovering a set of commonly used and mined process constraints. Therefore, we use the control-flow constraints defined in Declare \cite{pesic2006declarative}. Furthermore, we provide queries for discovering resource assignment constraints. Here, we provide queries to discover the resource allocation patterns of the well-known Workflow Resource Patterns \cite{russell2005workflow}. In addition to resource assignment there are cross-perspective patterns that relate to the control-flow and the resource perspectives at the same time.

The remainder of this paper is structured as follows: Section \ref{sec:background} describes the background and related work of the work at hand. Section \ref{sec:constraints} provides the actual SQL queries for discovering different constraints from event logs. Section \ref{sec:conclusion} concludes the paper.

\section{Background and Related Work} \label{sec:background}
Next, we describe the input data for our approach as well as fundamentals of declarative process modelling and automated discovery of models.

\subsection{Storing Event Log Data in Relational Databases} \label{sec:eventlog}
Our process mining approach takes as input \textit{(i) an event log}, i.e., a machine-recorded file that reports on the execution of activities during the enactment of the instances of a given process, and optionally \textit{(ii) organisational background knowledge}, i.e., prior knowledge about the roles, capabilities and the assignment of resources to organizational units.

In an event log, every process instance corresponds to a sequence (\textit{trace}) of recorded entries, namely, \textit{events}. We require that events contain an explicit reference to the enacted activity. For discovering resource-related aspects we additionally require an explicit reference to the operating resource. Both conditions are commonly contained in real-world event logs \cite{VanderAalst2011}.

The standardised, extensible storage format OpenXES was developed for the purpose of storing event data~\cite{Verbeek2011}. Lately, with \textit{RelationalXES (RXES)} a relational database architecture for storing event log data has been introduced \cite{relationalXES}. The RXES architecture uses a database to store the event log where traces and events are represented by tables with identifiers (IDs). RXES provides a full implementation of all OpenXES interfaces using the database as a backend. The database schema used in RXES allows for a significant reduction of redundancy by storing frequently occurring attributes only once rather than repeating them for every occurrence.

In this paper, we consider an event log to be available according to the RXES architecture and therefore, be stored in a conventional relational database. For readability we use a denormalized event log table, comprising only the attributes \textit{EventID} (unique identifier for each recorded event), \textit{TraceID} (unique identifier for the corresponding trace), \textit{ActivityID} (name of the corresponding activity the event refers to), \textit{Time} (date and time the event has occurred) as well as \textit{Identity} (identifier of the performing resource or person). In the remainder of the paper, we will use the shorthand notation $(a,i)$ for indicating an event of activity $a$ that has been executed by an identity $i$. The given events are ordered temporally so that timestamps are not encoded explicitly.

The example event log contains four traces comprising events of four different activities and executed by five different resources: \textit{$\langle$(a,$i_1$),(b,$i_1$),(c,$i_2$)$\rangle$}, \textit{$\langle$(b,$i_2$),(c,$i_2$)$\rangle$}, \textit{$\langle$(a,$i_2$),(d,$i_4$),(c,$i_2$)$\rangle$}, \textit{$\langle$(a,$i_5$),(a,$i_5$),(b,$i_1$),(c,$i_3$)$\rangle$}.

In the case that correlations between process execution and resource characteristics should be examined, organisational background information, e.g., in form of an organisational model must also be given in a relational database table. We build upon the generic organisational meta-model defined in \cite{bussler1998organisationsverwaltung}. A \textsf{Identity} represents an individual person that can be directly assigned to activities. A \textsf{Group} describes several resources as a whole. A \textsf{Relation} represents the interplay between \textsf{Identity} and \textsf{Group}. For \textsf{Relation}s, a \textsf{RelationType} specifies its interpretation. 
For instance, an organisation may have three \textsf{Group}s (Professor, Student, Admin), five \textsf{Identities} ($i_1$, $i_2$, $i_3$, $i_4$ and $i_5$) and the following four \textsf{Relation}s of \textsf{RelationType} ``role'' describing the roles that are assigned to each resource: \textit{($i_1$,role,Student)}, \textit{($i_2$,role,Professor)}, \textit{($i_3$,role,Professor)}, \textit{($i_4$,role,Admin)}, \textit{($i_5$,role,Student)}.


\subsection{Fundamentals of Declarative Process Mining} \label{sec:fundamentals}
In the following, we describe the basic aspects of declarative process modelling and the principle of automated discovery of such models from event log data.

\subsubsection{Declarative Process Mining}
During the last decade several declarative process modelling languages have been developed. For almost each declarative process modelling language a corresponding mining approach exists. The first approach to extract declarative constraints from event logs introduced by Maggi et al. is the \textit{DeclareMiner} \cite{Maggi2011}. Here, the user can select from a set of predefined Declare constraint templates the ones to be discovered. The system then generates all possible constraints by instantiating the chosen set of constraint templates with all possible combinations of occurring process elements provided in the event log. For example, the \textit{response} template consists of two placeholders for activities. Assuming that $|A|$ different activities occur in the event log, $|A|^2$ \textit{constraint candidates} are generated. All the resulting candidates are subsequently checked w.r.t. the event log. Additional mining parameters like PoE (Percentage of Events) or PoI (Percentage of Instances) are used to distinguish between valid and non-valid constraint candidates. Maggi et al. propose an evaluation of this algorithm with the adoption of a two-phase approach \cite{Maggi2012}. During the first phase, frequent sets of correlated activities are identified. The candidate constraints are only generated on the basis of these activities. In the second phase, the candidates are then checked in the same way as in \cite{Maggi2011}. Additionally, there are post-processing approaches that aim at simplifying the resulting Declare models in terms of
, a.o., redundancy elimination \cite{Maggi2013,CiccioM15} and disambiguation \cite{Bose2013}. In essence, the focus of these approaches is control flow with extensions to cover data \cite{Maggi20132}. The \textit{DPILMiner} \cite{schoenig2015bpmds} proposes a declarative mining approach to incorporate the resource perspective and to mine for a set of predefined resource assignment constraints.
All the mentioned approaches suffer performance issues w.r.t. real-life event logs. Furthermore, the set of constraint templates to be analysed is predefined, i.e., cannot be customised by the user.

Efficient algorithms to discover Declare models are presented in \cite{CiccioSLM15,Westergaard2013}. Westergaard presents the \textit{UnconstrainedMiner} \cite{Westergaard2013}, whose outstanding performance is obtained by constraint checking parallelisation and by relying on efficient data structures. The \textit{MINERful} approach \cite{CiccioM15} that has been implemented \cite{CiccioSLM15} in the ProM framework has shown to be the most efficient algorithm to discover control-flow constraints using the Declare language. Both approaches, however, are limited to discover control-flow constraints of the Declare language.

\subsubsection{Mining Metrics}
Checking constraint candidates provides for every constraint candidate the number of satisfactions in the event log. The constraint \textit{Response($a$,$b$)}, e.g., is satisfied three times in the example event log since in three cases of an occurrence of $a$, $b$ eventually follows in the same trace. Based on the number of satisfactions two metrics \textit{Support} and \textit{Confidence} are calculated that express the probability of a constraint to hold in the process. By comparing the values with user-defined thresholds, constraints are separated into non-valid and valid ones. In literature, there are two different definitions of support and confidence in the context of process mining. For our approach we adopt the most recent definition by Di Ciccio et al. \cite{CiccioM15} where the metrics are defined as follows:

\begin{itemize}
\item \textbf{Support:} It is the number of fulfilments of a constraint divided by the number of occurrences of the condition of a constraint. In the example log, the support of \textit{Response($a$,$b$)} is 0.75, as 3 $a$'s out of 4 fulfil the constraint. In case of constraints that do not depict implications like \textit{Existence} constraints in Declare, the Support is defined as the number of fulfilments divided by the number of traces in the log.

\item \textbf{Confidence:} It is the product of the support and the fraction of traces in the log where the condition (implications) or the constrained activity (not implications) occurs. The confidence of \textit{Response($a$,$b$)} is 0.75 $\cdot$  0.75 = 0.5625, since condition $a$ occurs in 3 traces out of 4.
\end{itemize}

The definitions above show that for discovering a certain constraint three different constraint specific values need to be extracted from event logs: \textit{(i) the number of occurrences of the condition of the constraint}, \textit{(ii) the number of fulfilments of the constraint}, and \textit{(iii) the fraction of traces in the log where the condition holds}.

\section{SQL Queries for Discovering Constraints} \label{sec:constraints}
In the following section we describe a set of different SQL queries for discovering constraints from event logs stored in a relational database table.

\subsection{Control Flow Constraints}
First, we give SQL queries for discovering control flow related constraints of the Declare language. Declare, a declarative process modeling language based on LTL (Linear Temporal Logic). Declare is characterized by a user-friendly graphical representation and formal semantics grounded in LTL. We define the queries of eight commonly known Declare constraint templates: \textit{Response}, \textit{AlternateReponse}, \textit{ChainResponse}, \textit{AlternatePrecedence}, \textit{Precedence}, \textit{ChainPrecedence}, \textit{RespondedExistence} and \textit{NotSuccession}. Note that all these templates have as parameters two activities. Hence, the result tables have the same number and type of columns. Consequently, all the queries can be executed together by connecting them with the SQL \textsf{UNION} operator.

Listing \ref{response} shows the query for discovering \textit{Response} constraints, i.e., ``if A occurs then eventually B occurs after A''.

{\small
\begin{lstlisting}[caption=SQL Query for discovering \textit{Response} constraints, label=response] 
SELECT 'response', TaskA, TaskB,
(CAST(COUNT(*) AS FLOAT)/CAST((SELECT COUNT(*) FROM Log WHERE Task LIKE TaskA) AS FLOAT)) AS Support,	
((CAST(COUNT(*) AS FLOAT)/CAST((SELECT COUNT(*) FROM Log WHERE Task LIKE TaskA) AS FLOAT)) * (CAST((SELECT COUNT(*) FROM (SELECT Instance FROM Log WHERE Task LIKE TaskA GROUP BY Instance)t2) AS FLOAT)/CAST((SELECT COUNT(*) FROM (SELECT Instance FROM Log GROUP BY Instance) t) AS FLOAT))) AS Confidence
FROM Log a, (SELECT a.Task AS TaskA, b.Task AS TaskB FROM Log a, Log b WHERE a.Task != b.Task GROUP BY a.Task, b.Task) x
WHERE a.Task = x.TaskA EXISTS (SELECT * FROM Log b WHERE b.Task = x.TaskB AND b.Instance = a.Instance AND b.Time] > a.Time])
GROUP BY x.TaskA, x.TaskB
HAVING (CAST(COUNT(*) AS FLOAT)/CAST((SELECT COUNT(*) FROM Log WHERE Task LIKE TaskA) AS FLOAT)) > 0.7 AND
 ((CAST(COUNT(*) AS FLOAT)/CAST((SELECT COUNT(*) FROM Log WHERE Task LIKE TaskA) AS FLOAT)) * (CAST((SELECT COUNT(*) FROM(SELECT Instance FROM Log WHERE Task LIKE TaskA GROUP BY Instance)t2) AS FLOAT) / CAST((SELECT COUNT(*) FROM(SELECT Instance FROM Log GROUP BY Instance) t) AS FLOAT))) > 0.5
\end{lstlisting}
}

Listing \ref{alternateresponse} shows the query for discovering \textit{AlternateResponse} constraints, i.e, ``if A occurs then eventually B occurs after A without other occurrences of A in between''.

{\small
\begin{lstlisting}[caption=SQL Query for discovering \textit{AlternateResponse} constraints, label=alternateresponse] 
SELECT 'alternateResponse', TaskA, TaskB,
(CAST(COUNT(*) AS FLOAT)/CAST((SELECT COUNT(*) FROM Log WHERE Task LIKE TaskA) AS FLOAT)) AS Support,	
((CAST(COUNT(*) AS FLOAT)/CAST((SELECT COUNT(*) FROM Log WHERE Task LIKE TaskA) AS FLOAT)) * (CAST((SELECT COUNT(*) FROM (SELECT Instance FROM Log WHERE Task LIKE TaskA GROUP BY Instance)t2) AS FLOAT)/CAST((SELECT COUNT(*) FROM (SELECT Instance FROM Log GROUP BY Instance) t) AS FLOAT))) AS Confidence
FROM Log a, (SELECT a.Task AS TaskA, b.Task AS TaskB FROM Log a, Log b WHERE a.Task != b.Task GROUP BY a.Task, b.Task) x
WHERE a.Task = x.TaskA AND EXISTS (SELECT * FROM Log b WHERE b.Task = x.TaskB AND b.Instance = a.Instance AND b.Time > a.Time)
 		AND NOT EXISTS(SELECT *  FROM Log b, Log c WHERE c.Instance = a.Instance AND c.Task = x.TaskA AND b.Instance = a.Instance AND b.Task = x.TaskB AND c.Time > a.Time AND c.Time < b.Time)
GROUP BY x.TaskA, x.TaskB
HAVING (CAST(COUNT(*) AS FLOAT)/CAST((SELECT COUNT(*) FROM Log WHERE Task LIKE TaskA) AS FLOAT)) > 0.7 AND
 ((CAST(COUNT(*) AS FLOAT)/CAST((SELECT COUNT(*) FROM Log WHERE Task LIKE TaskA) AS FLOAT)) * (CAST((SELECT COUNT(*) FROM(SELECT Instance FROM Log WHERE Task LIKE TaskA GROUP BY Instance)t2) AS FLOAT) / CAST((SELECT COUNT(*) FROM(SELECT Instance FROM Log GROUP BY Instance) t) AS FLOAT))) > 0.5
\end{lstlisting}
}

Listing \ref{chainresponse} shows the query for discovering \textit{ChainResponse} constraints, i.e., ``if A occurs then B occurs in the next position after A''

{\small
\begin{lstlisting}[caption=SQL Query for discovering \textit{ChainResponse} constraints, label=chainresponse] 
SELECT 'ChainResponse', TaskA, TaskB,
(CAST(COUNT(*) AS FLOAT)/CAST((SELECT COUNT(*) FROM Log WHERE Task LIKE TaskA) AS FLOAT)) AS Support,	
((CAST(COUNT(*) AS FLOAT)/CAST((SELECT COUNT(*) FROM Log WHERE Task LIKE TaskA) AS FLOAT)) * (CAST((SELECT COUNT(*) FROM (SELECT Instance FROM Log WHERE Task LIKE TaskA GROUP BY Instance)t2) AS FLOAT)/CAST((SELECT COUNT(*) FROM (SELECT Instance FROM Log GROUP BY Instance) t) AS FLOAT))) AS Confidence
FROM Log a, (SELECT a.Task AS TaskA, b.Task AS TaskB FROM Log a, Log b WHERE a.Task != b.Task GROUP BY a.Task, b.Task) x
WHERE a.Task = x.TaskA AND EXISTS (SELECT * FROM Log b WHERE b.Task = x.TaskB AND b.Instance = a.Instance AND b.Time > a.Time)
 		AND NOT EXISTS(SELECT * FROM Log b, Log c WHERE c.Instance = a.Instance AND b.Instance = a.Instance AND b.Task = x.TaskB AND c.Time > a.Time AND c.Time < b.Time)
GROUP BY x.TaskA, x.TaskB
HAVING (CAST(COUNT(*) AS FLOAT)/CAST((SELECT COUNT(*) FROM Log WHERE Task LIKE TaskA) AS FLOAT)) > 0.7 AND
 ((CAST(COUNT(*) AS FLOAT)/CAST((SELECT COUNT(*) FROM Log WHERE Task LIKE TaskA) AS FLOAT)) * (CAST((SELECT COUNT(*) FROM(SELECT Instance FROM Log WHERE Task LIKE TaskA GROUP BY Instance)t2) AS FLOAT) / CAST((SELECT COUNT(*) FROM(SELECT Instance FROM Log GROUP BY Instance) t) AS FLOAT))) > 0.5
\end{lstlisting}
}

Listing \ref{precedence} shows the query for discovering \textit{Precedence} constraints, i.e., ``if B occurs then A occurs before B''.

{\small
\begin{lstlisting}[caption=SQL Query for discovering \textit{Precedence} constraints, label=precedence] 
SELECT 'Precedence', TaskA, TaskB,
(CAST(COUNT(*) AS FLOAT)/CAST((SELECT COUNT(*) FROM Log WHERE Task LIKE TaskB) AS FLOAT)) AS Support,	
((CAST(COUNT(*) AS FLOAT)/CAST((SELECT COUNT(*) FROM Log WHERE Task LIKE TaskB) AS FLOAT)) * (CAST((SELECT COUNT(*) FROM (SELECT Instance FROM Log WHERE Task LIKE TaskB GROUP BY Instance)t2) AS FLOAT)/CAST((SELECT COUNT(*) FROM (SELECT Instance FROM Log GROUP BY Instance) t) AS FLOAT))) AS Confidence
FROM Log a, (SELECT a.Task AS TaskA, b.Task AS TaskB FROM Log a, Log b WHERE a.Task != b.Task GROUP BY a.Task, b.Task) x
WHERE a.Task = x.TaskB AND EXISTS (SELECT * FROM Log b WHERE b.Task = x.TaskA AND b.Instance = a.Instance AND b.Time < a.Time)
GROUP BY x.TaskA, x.TaskB
HAVING (CAST(COUNT(*) AS FLOAT)/CAST((SELECT COUNT(*) FROM Log WHERE Task LIKE TaskB) AS FLOAT)) > 0.7 AND
 ((CAST(COUNT(*) AS FLOAT)/CAST((SELECT COUNT(*) FROM Log WHERE Task LIKE TaskB) AS FLOAT)) * (CAST((SELECT COUNT(*) FROM(SELECT Instance FROM Log WHERE Task LIKE TaskB GROUP BY Instance)t2) AS FLOAT) / CAST((SELECT COUNT(*) FROM(SELECT Instance FROM Log GROUP BY Instance) t) AS FLOAT))) > 0.5
\end{lstlisting}
}

Listing \ref{alternateprecedence} shows the query for discovering \textit{AlternatePrecedence} constraints, i.e., ``if B occurs then A occurs before B without other occurrences of B in between''.

{\small
\begin{lstlisting}[caption=SQL Query for discovering \textit{AlternatePrecedence} constraints, label=alternateprecedence] 
SELECT 'alternatePrecedence', TaskA, TaskB,
(CAST(COUNT(*) AS FLOAT)/CAST((SELECT COUNT(*) FROM Log WHERE Task LIKE TaskB) AS FLOAT)) AS Support,	
((CAST(COUNT(*) AS FLOAT)/CAST((SELECT COUNT(*) FROM Log WHERE Task LIKE TaskB) AS FLOAT)) * (CAST((SELECT COUNT(*) FROM (SELECT Instance FROM Log WHERE Task LIKE TaskB GROUP BY Instance)t2) AS FLOAT)/CAST((SELECT COUNT(*) FROM (SELECT Instance FROM Log GROUP BY Instance) t) AS FLOAT))) AS Confidence
FROM Log a, (SELECT a.Task AS TaskA, b.Task AS TaskB FROM Log a, Log b WHERE a.Task != b.Task GROUP BY a.Task, b.Task) x
WHERE a.Task = x.TaskB AND EXISTS (SELECT * FROM Log b WHERE b.Task = x.TaskA AND b.Instance = a.Instance AND b.Time < a.Time)
		AND NOT EXISTS(SELECT *  FROM Log b, Log c WHERE c.Instance = a.Instance AND c.Task = x.TaskB AND b.Instance = a.Instance AND b.Task = x.TaskA AND c.Time < a.Time AND c.Time > b.Time)
GROUP BY x.TaskA, x.TaskB
HAVING (CAST(COUNT(*) AS FLOAT)/CAST((SELECT COUNT(*) FROM Log WHERE Task LIKE TaskB) AS FLOAT)) > 0.7 AND
 ((CAST(COUNT(*) AS FLOAT)/CAST((SELECT COUNT(*) FROM Log WHERE Task LIKE TaskB) AS FLOAT)) * (CAST((SELECT COUNT(*) FROM(SELECT Instance FROM Log WHERE Task LIKE TaskB GROUP BY Instance)t2) AS FLOAT) / CAST((SELECT COUNT(*) FROM(SELECT Instance FROM Log GROUP BY Instance) t) AS FLOAT))) > 0.5
\end{lstlisting}
}

Listing \ref{chainprecedence} shows the query for discovering \textit{ChainPrecedence} constraints, i.e., ``if B occurs then A occurs in the next position before B''.

{\small
\begin{lstlisting}[caption=SQL Query for discovering \textit{ChainPrecedence} constraints, label=chainprecedence] 
SELECT 'chainPrecedence', TaskA, TaskB,
(CAST(COUNT(*) AS FLOAT)/CAST((SELECT COUNT(*) FROM Log WHERE Task LIKE TaskB) AS FLOAT)) AS Support,	
((CAST(COUNT(*) AS FLOAT)/CAST((SELECT COUNT(*) FROM Log WHERE Task LIKE TaskB) AS FLOAT)) * (CAST((SELECT COUNT(*) FROM (SELECT Instance FROM Log WHERE Task LIKE TaskB GROUP BY Instance)t2) AS FLOAT)/CAST((SELECT COUNT(*) FROM (SELECT Instance FROM Log GROUP BY Instance) t) AS FLOAT))) AS Confidence
FROM Log a, (SELECT a.Task AS TaskA, b.Task AS TaskB FROM Log a, Log b WHERE a.Task != b.Task GROUP BY a.Task, b.Task) x
WHERE a.Task = x.TaskB AND EXISTS (SELECT * FROM Log b WHERE b.Task = x.TaskA AND b.Instance = a.Instance AND b.Time < a.Time)
		AND NOT EXISTS(SELECT *  FROM Log b, Log c WHERE c.Instance = a.Instance AND b.Instance = a.Instance AND b.Task = x.TaskA AND c.Time < a.Time AND c.Time > b.Time)
GROUP BY x.TaskA, x.TaskB
HAVING (CAST(COUNT(*) AS FLOAT)/CAST((SELECT COUNT(*) FROM Log WHERE Task LIKE TaskB) AS FLOAT)) > 0.7 AND
 ((CAST(COUNT(*) AS FLOAT)/CAST((SELECT COUNT(*) FROM Log WHERE Task LIKE TaskB) AS FLOAT)) * (CAST((SELECT COUNT(*) FROM(SELECT Instance FROM Log WHERE Task LIKE TaskB GROUP BY Instance)t2) AS FLOAT) / CAST((SELECT COUNT(*) FROM(SELECT Instance FROM Log GROUP BY Instance) t) AS FLOAT))) > 0.5
\end{lstlisting}
}

Listing \ref{respondedexistence} shows the query for discovering \textit{RespondedExistence} constraints, i.e., ``if A occurs then B occurs before or after A''.

{\small
\begin{lstlisting}[caption=SQL Query for discovering \textit{RespondedExistence} constraints, label=respondedexistence] 
SELECT 'respondedExistence', TaskA, TaskB,
(CAST(COUNT(*) AS FLOAT)/CAST((SELECT COUNT(*) FROM Log WHERE Task LIKE TaskA) AS FLOAT)) AS Support,	
((CAST(COUNT(*) AS FLOAT)/CAST((SELECT COUNT(*) FROM Log WHERE Task LIKE TaskA) AS FLOAT)) * (CAST((SELECT COUNT(*) FROM (SELECT Instance FROM Log WHERE Task LIKE TaskA GROUP BY Instance)t2) AS FLOAT)/CAST((SELECT COUNT(*) FROM (SELECT Instance FROM Log GROUP BY Instance) t) AS FLOAT))) AS Confidence
FROM Log a, (SELECT a.Task AS TaskA, b.Task AS TaskB FROM Log a, Log b WHERE a.Task != b.Task GROUP BY a.Task, b.Task) x
WHERE a.Task = x.TaskB AND EXISTS (SELECT * FROM Log b WHERE b.Task = x.TaskA AND b.Instance = a.Instance)
GROUP BY x.TaskA, x.TaskB
HAVING (CAST(COUNT(*) AS FLOAT)/CAST((SELECT COUNT(*) FROM Log WHERE Task LIKE TaskA) AS FLOAT)) > 0.7 AND
 ((CAST(COUNT(*) AS FLOAT)/CAST((SELECT COUNT(*) FROM Log WHERE Task LIKE TaskA) AS FLOAT)) * (CAST((SELECT COUNT(*) FROM(SELECT Instance FROM Log WHERE Task LIKE TaskA GROUP BY Instance)t2) AS FLOAT) / CAST((SELECT COUNT(*) FROM(SELECT Instance FROM Log GROUP BY Instance) t) AS FLOAT))) > 0.5
\end{lstlisting}
}

Listing \ref{notsuccession} shows the query for discovering \textit{NotSuccession} constraints, i.e., ``if A occurs then B cannot eventually occur after A''.

{\small
\begin{lstlisting}[caption=SQL Query for discovering \textit{NotSuccession} constraints, label=notsuccession] 
SELECT 'notSuccession', TaskA, TaskB,
(CAST(COUNT(*) AS FLOAT)/CAST((SELECT COUNT(*) FROM Log WHERE Task LIKE TaskA) AS FLOAT)) AS Support,	
((CAST(COUNT(*) AS FLOAT)/CAST((SELECT COUNT(*) FROM Log WHERE Task LIKE TaskA) AS FLOAT)) * (CAST((SELECT COUNT(*) FROM (SELECT Instance FROM Log WHERE Task LIKE TaskA GROUP BY Instance)t2) AS FLOAT)/CAST((SELECT COUNT(*) FROM (SELECT Instance FROM Log GROUP BY Instance) t) AS FLOAT))) AS Confidence
FROM Log a, (SELECT a.Task AS TaskA, b.Task AS TaskB FROM Log a, Log b WHERE a.Task != b.Task GROUP BY a.Task, b.Task) x
WHERE a.Task = x.TaskB AND a.Time < ALL (SELECT Time FROM Log b WHERE b.Task = x.TaskA AND b.Instance = a.Instance)
 		AND EXISTS (SELECT * FROM Log b WHERE b.Task = x.TaskA AND b.Instance = a.Instance)
		AND a.Time > ALL(SELECT Time FROM Log b WHERE b.Task = x.TaskB AND b.Instance = a.Instance)
GROUP BY x.TaskA, x.TaskB
HAVING (CAST(COUNT(*) AS FLOAT)/CAST((SELECT COUNT(*) FROM Log WHERE Task LIKE TaskA) AS FLOAT)) > 0.7 AND
 ((CAST(COUNT(*) AS FLOAT)/CAST((SELECT COUNT(*) FROM Log WHERE Task LIKE TaskA) AS FLOAT)) * (CAST((SELECT COUNT(*) FROM(SELECT Instance FROM Log WHERE Task LIKE TaskA GROUP BY Instance)t2) AS FLOAT) / CAST((SELECT COUNT(*) FROM(SELECT Instance FROM Log GROUP BY Instance) t) AS FLOAT))) > 0.5
\end{lstlisting}
}

\subsection{Resource Assignment Constraints}
The organisational perspective of processes is extensively discussed by the workflow resource patterns. These patterns capture the various ways in which resources are represented and utilised in processes \cite{russell_workflow_2005}. Of specific interest to process mining are the creation patterns since they capture options to configure the different ways in which resources can be assigned to activities at the level of a process model. Among them, the patterns that we discuss include:
\textit{Direct Distribution} is the ability to specify at design time the identity of the resource that will execute a task.
\textit{Role-Based Distribution} is the ability to specify at design time that a task can only be executed by resources that correspond to a given role.
\textit{Organisational Distribution} is the ability to offer or allocate activity instances to resources based on their organisational position and their relationship with other resources.
\textit{Separation of duties} is the ability to specify that two tasks must be allocated to different resources in a given process instance.
\textit{Retain Familiar} is the ability to allocate an activity instance within a given process instance to the same resource that performed a preceding activity instance. This pattern is also known as \textit{binding of duties}.

Listing \ref{directAllocation} shows the query for discovering \textit{DirectAllocation} constraints, i.e., ``if A occurs then it is executed by person P''.

{\small
\begin{lstlisting}[caption=SQL Query for discovering \textit{DirectAllocation} constraints, label=directAllocation] 
SELECT 'DirectAllocation', TaskA, a.Resource,
(CAST(COUNT(*) AS FLOAT)/CAST((SELECT COUNT(*) FROM Log WHERE Task LIKE TaskA) AS FLOAT)) AS Support,	
((CAST(COUNT(*) AS FLOAT)/CAST((SELECT COUNT(*) FROM Log WHERE Task LIKE TaskA) AS FLOAT)) * (CAST((SELECT COUNT(*) FROM (SELECT Instance FROM Log WHERE Task LIKE TaskA GROUP BY Instance)t2) AS FLOAT)/CAST((SELECT COUNT(*) FROM (SELECT Instance FROM Log GROUP BY Instance) t) AS FLOAT))) AS Confidence
FROM Log a, (SELECT a.Task AS TaskA, b.Task AS TaskB FROM Log a, Log b WHERE a.Task != b.Task GROUP BY a.Task, b.Task) x
WHERE a.Task = x.TaskA
GROUP BY x.TaskA, a.Resource
HAVING (CAST(COUNT(*) AS FLOAT)/CAST((SELECT COUNT(*) FROM Log WHERE Task LIKE TaskA) AS FLOAT)) > 0.7 AND
 ((CAST(COUNT(*) AS FLOAT)/CAST((SELECT COUNT(*) FROM Log WHERE Task LIKE TaskA) AS FLOAT)) * (CAST((SELECT COUNT(*) FROM(SELECT Instance FROM Log WHERE Task LIKE TaskA GROUP BY Instance)t2) AS FLOAT) / CAST((SELECT COUNT(*) FROM(SELECT Instance FROM Log GROUP BY Instance) t) AS FLOAT))) > 0.5
\end{lstlisting}
}

Listing \ref{roleBasedAllocation} shows the query for discovering \textit{RoleBasedAllocation} constraints, i.e., ``if A occurs then it is executed by person in role R''.

{\small
\begin{lstlisting}[caption=SQL Query for discovering \textit{RoleBasedAllocation} constraints, label=roleBasedAllocation] 
SELECT  'RoleBasedAllocation', TaskA, r1.Group,
(CAST(COUNT(*) AS FLOAT)/CAST((SELECT COUNT(*) FROM Log WHERE Task LIKE TaskA) AS FLOAT)) AS Support,	
((CAST(COUNT(*) AS FLOAT)/CAST((SELECT COUNT(*) FROM Log WHERE Task LIKE TaskA) AS FLOAT)) * (CAST((SELECT COUNT(*) FROM (SELECT Instance FROM Log WHERE Task LIKE TaskA GROUP BY Instance)t2) AS FLOAT)/CAST((SELECT COUNT(*) FROM (SELECT Instance FROM Log GROUP BY Instance) t) AS FLOAT))) AS Confidence
FROM Log a, Relation r1, (SELECT a.Task AS TaskA, b.Task AS TaskB FROM Log a, Log b WHERE a.Task != b.Task GROUP BY a.Task, b.Task) x
WHERE a.Task = x.TaskA AND r1.RelationType LIKE 'role' AND a.Resource = r1.Resource
HAVING (CAST(COUNT(*) AS FLOAT)/CAST((SELECT COUNT(*) FROM Log WHERE Task LIKE TaskA) AS FLOAT)) > 0.7 AND
	((CAST(COUNT(*) AS FLOAT)/CAST((SELECT COUNT(*) FROM Log WHERE Task LIKE TaskA) AS FLOAT)) * (CAST((SELECT COUNT(*) FROM(SELECT Instance FROM Log WHERE Task LIKE TaskA GROUP BY Instance)t2) AS FLOAT) / CAST((SELECT COUNT(*) FROM(SELECT Instance FROM Log GROUP BY Instance) t) AS FLOAT))) > 0.5
\end{lstlisting}
}

Listing \ref{BindingOfDuties} shows the query for discovering \textit{BindingOfDuties} constraints, i.e., ``A and B are performed by the same person''.

{\small
\begin{lstlisting}[caption=SQL Query for discovering \textit{BindingOfDuties} constraints, label=BindingOfDuties] 
SELECT  'BindingOfDuties', TaskA, TaskB,
(CAST(COUNT(*) AS FLOAT)/CAST((SELECT COUNT(*) FROM Log a WHERE a.Task LIKE TaskA AND EXISTS(SELECT * FROM Log b WHERE b.Task = x.TaskB AND b.Instance = a.Instance)) AS FLOAT)) AS Support,
((CAST(COUNT(*) AS FLOAT)/CAST((SELECT COUNT(*) FROM Log a WHERE a.Task LIKE TaskA AND EXISTS(SELECT * FROM Log b WHERE b.Task = x.TaskB AND b.Instance = a.Instance)) AS FLOAT)) * (CAST((SELECT COUNT(*) FROM(SELECT Instance FROM Log a WHERE a.Task LIKE TaskA AND EXISTS(SELECT *  FROM Log b WHERE b.Task = x.TaskB AND b.Instance = a.Instance) GROUP BY Instance)t2) AS FLOAT)/CAST((SELECT COUNT(*) FROM(SELECT Instance FROM Log GROUP BY Instance) t) AS FLOAT))) AS Confidence
FROM Log a, (SELECT a.Task AS TaskA, b.Task AS TaskB FROM Log a, Log b WHERE a.Task != b.Task GROUP BY a.Task, b.Task) x
WHERE a.Task = x.TaskA AND EXISTS (SELECT * FROM Log b WHERE b.Task = x.TaskB AND b.Instance = a.Instance)
	AND a.Resource = ALL (SELECT b.Resource FROM Log b WHERE b.Task = x.TaskB AND b.Instance = a.Instance)
HAVING (CAST(COUNT(*) AS FLOAT)/CAST((SELECT COUNT(*) FROM Log a WHERE a.Task LIKE TaskA AND EXISTS(SELECT * FROM Log b WHERE b.Task = x.TaskB AND b.Instance = a.Instance)) AS FLOAT)) > 0.7 AND
	((CAST(COUNT(*) AS FLOAT)/CAST((SELECT COUNT(*) FROM Log a WHERE a.Task LIKE TaskA AND EXISTS(SELECT * FROM Log b WHERE b.Task = x.TaskB AND b.Instance = a.Instance)) AS FLOAT)) * (CAST((SELECT COUNT(*) FROM(SELECT Instance FROM Log a WHERE a.Task LIKE TaskA AND EXISTS(SELECT *  FROM Log b WHERE b.Task = x.TaskB AND b.Instance = a.Instance) GROUP BY Instance)t2) AS FLOAT)/CAST((SELECT COUNT(*) FROM(SELECT Instance FROM Log GROUP BY Instance) t) AS FLOAT))) > 0.5
\end{lstlisting}
}

Listing \ref{SeparationOfDuties} shows the query for discovering \textit{SeparationOfDuties} constraints, i.e., ``A and B are performed by different persons''.

{\small
\begin{lstlisting}[caption=SQL Query for discovering \textit{SeparationOfDuties} constraints, label=SeparationOfDuties] 
SELECT  'SeparationOfDuties', TaskA, TaskB,
(CAST(COUNT(*) AS FLOAT)/CAST((SELECT COUNT(*) FROM Log a WHERE a.Task LIKE TaskA AND EXISTS(SELECT * FROM Log b WHERE b.Task = x.TaskB AND b.Instance = a.Instance)) AS FLOAT)) AS Support,
((CAST(COUNT(*) AS FLOAT)/CAST((SELECT COUNT(*) FROM Log a WHERE a.Task LIKE TaskA AND EXISTS(SELECT * FROM Log b WHERE b.Task = x.TaskB AND b.Instance = a.Instance)) AS FLOAT)) * (CAST((SELECT COUNT(*) FROM(SELECT Instance FROM Log a WHERE a.Task LIKE TaskA AND EXISTS(SELECT *  FROM Log b WHERE b.Task = x.TaskB AND b.Instance = a.Instance) GROUP BY Instance)t2) AS FLOAT)/CAST((SELECT COUNT(*) FROM(SELECT Instance FROM Log GROUP BY Instance) t) AS FLOAT))) AS Confidence
FROM Log a, (SELECT a.Task AS TaskA, b.Task AS TaskB FROM Log a, Log b WHERE a.Task != b.Task GROUP BY a.Task, b.Task) x
WHERE a.Task = x.TaskA AND EXISTS (SELECT * FROM Log b WHERE b.Task = x.TaskB AND b.Instance = a.Instance)
	AND a.Resource != ALL (SELECT b.Resource FROM Log b WHERE b.Task = x.TaskB AND b.Instance = a.Instance)
HAVING (CAST(COUNT(*) AS FLOAT)/CAST((SELECT COUNT(*) FROM Log a WHERE a.Task LIKE TaskA AND EXISTS(SELECT * FROM Log b WHERE b.Task = x.TaskB AND b.Instance = a.Instance)) AS FLOAT)) > 0.7 AND
	((CAST(COUNT(*) AS FLOAT)/CAST((SELECT COUNT(*) FROM Log a WHERE a.Task LIKE TaskA AND EXISTS(SELECT * FROM Log b WHERE b.Task = x.TaskB AND b.Instance = a.Instance)) AS FLOAT)) * (CAST((SELECT COUNT(*) FROM(SELECT Instance FROM Log a WHERE a.Task LIKE TaskA AND EXISTS(SELECT *  FROM Log b WHERE b.Task = x.TaskB AND b.Instance = a.Instance) GROUP BY Instance)t2) AS FLOAT)/CAST((SELECT COUNT(*) FROM(SELECT Instance FROM Log GROUP BY Instance) t) AS FLOAT))) > 0.5
\end{lstlisting}
}

\subsection{Cross-Perspective Constraints}
In addition to resource assignment there are cross-perspective patterns that relate to the control-flow and the organisational perspectives at the same time. They can be found in different application areas, e.g., in cases where the execution of certain process steps is bound to conditions that apply only for certain resources. These patterns include the following ones described in~\cite{schoenig2015bpmds}:
\textit{RoleBasedResponse} represents situations in which an activity must be executed after another one for specific organisational roles but not for others. For instance, resources with role \textit{Student} must always request for permission to book a work trip before buying the transport tickets; this might not be required for the other roles, though. Applying the same reasoning, \textit{RoleBasedPrecedence} represents situations in which for specific organisational roles an activity must have been executed in order to perform another task. Further cross-perspective patterns could be defined following a similar approach.

Listing \ref{RoleBasedResponse} shows the query for discovering \textit{RoleBasedResponse} constraints, i.e., ``if A occurs and is performed by a person in role R then eventually B occurs after A''.

{\small
\begin{lstlisting}[caption=SQL Query for discovering \textit{RoleBasedResponse} constraints, label=RoleBasedResponse] 
SELECT 'RoleBasedResponse', TaskA, TaskB, Group
(CAST(COUNT(*) AS FLOAT)/CAST((SELECT COUNT(*) FROM Log, Relation WHERE RelationType LIKE 'role' AND Log.Resource = Relation.Resource AND Task LIKE TaskA AND Relation.Group LIKE r1.Group) AS FLOAT)) AS Support,
	((CAST(COUNT(*) AS FLOAT)/CAST((SELECT COUNT(*) FROM Log, Relation WHERE RelationType LIKE 'role' AND Log.Resource = Relation.Resource AND Task LIKE TaskA AND Relation.Group LIKE r1.Group) AS FLOAT)) * (CAST((SELECT COUNT(*) FROM(SELECT Instance FROM Log, Relation WHERE RelationType LIKE 'role' AND Log.Resource = Relation.Resource AND Task LIKE TaskA AND Relation.Group LIKE r1.Group GROUP BY Instance)t2) AS FLOAT)/CAST((SELECT COUNT(*) FROM(SELECT InstanceFROM Log GROUP BY Instance) t) AS FLOAT))) AS Confidence
FROM Log a, Relation r1, (SELECT a.Task AS TaskA, b.Task AS TaskB FROM Log a, Log b WHERE a.Task != b.Task GROUP BY a.Task, b.Task) x
WHERE a.Task = x.TaskA AND r1.RelationType LIKE 'role' AND a.Resource = r1.Resource
	AND a.Time < ALL (SELECT Time FROM Log b WHERE b.Task = x.TaskB AND b.Instance = a.Instance)
	AND EXISTS (SELECT * FROM Log b WHERE b.Task = x.TaskB AND b.Instance = a.Instance)
GROUP BY x.TaskA, x.TaskB, r1.Group
HAVING (CAST(COUNT(*) AS FLOAT)/CAST((SELECT COUNT(*) FROM Log, Relation WHERE RelationType LIKE 'role' AND Log.Resource = Relation.Resource AND Task LIKE TaskA AND Relation.Group LIKE r1.Group) AS FLOAT)) > 0.7
	AND ((CAST(COUNT(*) AS FLOAT)/CAST((SELECT COUNT(*) FROM Log, Relation WHERE RelationType LIKE 'role' AND Log.Resource = Relation.Resource AND Task LIKE TaskA AND Relation.Group LIKE r1.Group) AS FLOAT)) * (CAST((SELECT COUNT(*) FROM(SELECT Instance FROM Log, Relation WHERE RelationType LIKE 'role' AND Log.Resource = Relation.Resource AND Task LIKE TaskA AND Relation.Group LIKE r1.Group GROUP BY Instance)t2) AS FLOAT)/CAST((SELECT COUNT(*) FROM(SELECT Instance FROM Log GROUP BY Instance) t) AS FLOAT))) > 0.5
\end{lstlisting}
}

Listing \ref{RoleBasedPrecedence} shows the query for discovering \textit{RoleBasedPrecedence} constraints, i.e., ``if B occurs and is performed by a person in role R, then A occurs before B''.

{\small
\begin{lstlisting}[caption=SQL Query for discovering \textit{RoleBasedPrecedence} constraints, label=RoleBasedPrecedence] 
SELECT 'RoleBasedPrecedence', TaskA, TaskB, Group
(CAST(COUNT(*) AS FLOAT)/CAST((SELECT COUNT(*) FROM Log, Relation WHERE RelationType LIKE 'role' AND Log.Resource = Relation.Resource AND Task LIKE TaskB AND Relation.Group LIKE r1.Group) AS FLOAT)) AS Support,
	((CAST(COUNT(*) AS FLOAT)/CAST((SELECT COUNT(*) FROM Log, Relation WHERE RelationType LIKE 'role' AND Log.Resource = Relation.Resource AND Task LIKE TaskB AND Relation.Group LIKE r1.Group) AS FLOAT)) * (CAST((SELECT COUNT(*) FROM(SELECT Instance FROM Log, Relation WHERE RelationType LIKE 'role' AND Log.Resource = Relation.Resource AND Task LIKE TaskB AND Relation.Group LIKE r1.Group GROUP BY Instance)t2) AS FLOAT)/CAST((SELECT COUNT(*) FROM(SELECT InstanceFROM Log GROUP BY Instance) t) AS FLOAT))) AS Confidence
FROM Log a, Relation r1, (SELECT a.Task AS TaskA, b.Task AS TaskB FROM Log a, Log b WHERE a.Task != b.Task GROUP BY a.Task, b.Task) x
WHERE a.Task = x.TaskB AND r1.RelationType LIKE 'role' AND a.Resource = r1.Resource
	AND a.Time < ALL (SELECT Time FROM Log b WHERE b.Task = x.TaskA AND b.Instance = a.Instance)
	AND EXISTS (SELECT * FROM Log b WHERE b.Task = x.TaskA AND b.Instance = a.Instance)
GROUP BY x.TaskA, x.TaskB, r1.Group
HAVING (CAST(COUNT(*) AS FLOAT)/CAST((SELECT COUNT(*) FROM Log, Relation WHERE RelationType LIKE 'role' AND Log.Resource = Relation.Resource AND Task LIKE TaskB AND Relation.Group LIKE r1.Group) AS FLOAT)) > 0.7
	AND ((CAST(COUNT(*) AS FLOAT)/CAST((SELECT COUNT(*) FROM Log, Relation WHERE RelationType LIKE 'role' AND Log.Resource = Relation.Resource AND Task LIKE TaskB AND Relation.Group LIKE r1.Group) AS FLOAT)) * (CAST((SELECT COUNT(*) FROM(SELECT Instance FROM Log, Relation WHERE RelationType LIKE 'role' AND Log.Resource = Relation.Resource AND Task LIKE TaskB AND Relation.Group LIKE r1.Group GROUP BY Instance)t2) AS FLOAT)/CAST((SELECT COUNT(*) FROM(SELECT Instance FROM Log GROUP BY Instance) t) AS FLOAT))) > 0.5
\end{lstlisting}
}

\section{Conlusion} \label{sec:conclusion}
In this paper, we introduced an SQL-based declarative process mining approach that analyses event log data stored in relational databases. While existing declarative process mining approaches either suffer performance issues with real-life event logs or limit their search space to a specific and fixed set of constraints, the SQLMiner approach constitutes a trade-off between performance and customisation capabilities. Leveraging relational database performance technology the approach is fast even without limiting itself to certain predefined set of constraints. Queries can be customised and comprise process perspectives apart from control flow, such as organisational aspects.

\glsresetall



\bibliographystyle{ieeetr}
\bibliography{paper}

\end{document}